# Photonic Structures to Achieve High-Performance Dew-Harvesting in a 24-h Day-Night Cycle

ZHANG Zheng[1], DONG Lining[2], DONG Minghao[1], RUAN Shiting[2], & CHEN Zhen[1*]

[1] *Jiangsu Key Laboratory for Design and Manufacturing of Precision Medicine Equipment, School of Mechanical Engineering, Southeast University, Nanjing 210096, China*
[2] *Shanghai Institute of Satellite Engineering, Shanghai 201109, China*

*\* To whom correspondence should be addressed: zhenchen@seu.edu.cn*

**Abstract**

Although most prior research on the dew-harvesting technology has focused on nocturnal operations, achieving round-the-clock freshwater harvesting remains crucial. However, daytime dew-harvesting faces two key challenges as compared to its nighttime counterpart: the high solar irradiance and the large contrast between the ambient temperature and the dewpoint. To address these challenges and guide the photonic design, we develop a theoretical framework to analyze dew-harvesting in a 24-h day-night cycle. Using Nanjing as an example, our analyses reveal that, in the solar regime, a minimum average solar reflectivity of 0.92 is required; in the infrared regime, a 10% reduction in absorptivity outside the 8-13 μm transparency window is equivalent to a 5.9% enhancement in emissivity within the window. Guided by these findings, we propose a photonic design, which, in a synthetic experiment with measured meteorological datasets, achieves a water production rate of 313 g/m²-day, in which nearly 40% is contributed by daytime. This performance reaches approximately 70% of the theoretical maximum predicted using the ideal spectrum. We end by optimizing the layout of condensers in practical applications.

**Key words**: daytime dew-harvesting; radiative cooling; photonic design

# 1 Introduction

Atmospheric vapor presents a promising opportunity to increase fresh water sources by approximately 15% [1], offering crucial support in mitigating the global water scarcity. Many efforts, including fog collection [2-5] and the adsorption-based water harvesting technology [2,6-13], have been made to utilize this resource. In addition, desalination [2,14-20] converts salty seawater to fresh water. Despite their potential, fog collection and desalination are limited to specific climates or locations. Besides, the requirement of additional energy inputs and the cost of adsorption materials challenge the scalability of adsorption-based technology, especially in underdeveloped regions.

Dew-harvesting technology uses radiative cooling [21-28] to cool a condenser below the dewpoint, thereby condensing the atmospheric water vapor in the vicinity into liquid form [29]. Compared to fog collection and desalination technologies, dew-harvesting is not restricted to specific climatic or geographical conditions. Moreover, it does not require complex infrastructure and additional energy input, making it particularly suitable for water-scarce inland regions.

Dew-harvesting at nighttime has a long history [30-38]. It is documented that the ancient Greeks used this technology for water supply to the city of Theodosia [39]. Most previous studies rely on natural materials that have weak intrinsic selectivity in their spectra [30-37]. Dong, Chen, and co-workers point out the importance of the photonic design of condensers to fully utilize the potential of outer space [38]. For instance, by incorporating an ideal selective emitter, which has unit (zero) emissivity inside (outside) the 8-13 μm transparency window, dew-harvesting technology can be effectively applied in arid regions with low relative humidity (e.g. $RH < 40\%$) [38]. This is different to the optimal spectrum for desalination [2,14-20], in which an infrared (IR) blackbody is required for hot steam condensation.

With the desire of providing round-the-clock fresh water passively, dew-harvesting at daytime has recently attracted broad interest. Previous studies have achieved daytime dew-harvesting using photonic designs with high solar reflectivity [40,41], since, in principle, daytime could produce nearly as much dew as possible as nighttime, as long as the condenser is able to reflect all the sunlight. However, these works achieved only a small amount of water yield under weak light conditions, which do not harness the full potential of daytime dew-harvesting. While these designs emphasize the high solar reflectivity, they ignore the importance of the IR selectivity, as highlighted in Ref. [38]. Additionally, some experiments pre-humidified the ambient air before supplying it to the condenser [40], resulting in inconsistency between the implemented configurations and the theoretical models with which they are compared [38]. This inconsistency leads to significant underestimate of the potential of the corresponding experimental configuration.

In this work, we extend our previous theoretical framework of dew-harvesting from nighttime to daytime. Using Nanjing, a typical city in the east coast of China with humid climates, as a concrete example, this framework indicates that an average solar reflectivity of at least 0.92 is required for the effective 24-hour dew-harvesting. In the IR regime, a 10% reduction of the absorptivity outside the transparency window (8-13 μm) is equivalent to a 5.9% increase of the emissivity within the window, highlighting the importance of a selective spectrum in the IR regime.

To validate our findings, we present a multilayer photonic design featuring high solar reflectivity and selective emissivity in the IR regime. This design enables continuous 24-hour dew-harvesting in a virtual experiment, and achieves a water production rate of 313 g/m²-day, in which nearly 40% is contributed by daytime. This performance reaches approximately 70% of the theoretical limit with the ideal spectrum. We also highlight the distinction between the global and the local humidity in experiments where the ambient air is pre-humified before supplying to the condenser, and suggest a distributed instead of a centralized layout of the condenser panels to maximize the water production yield.

## 2. Framework

Figure 1a illustrates a setup for daytime dew-harvesting. The core component is a photonic design that serves as the condenser. Above the condenser, a chamber enclosed by an IR-transparent window and thermal insulating materials minimize the parasitic heat convection, characterized with a convection coefficient, $h_{parastic}$, with SI units of W/m²-K. Dew condenses on the exposed bottom surface, which avoids suppression of radiative cooling and evaporation induced by sunlight [42]. Below the condenser, the convection with a different coefficient, $h$, supplies fresh air flow continuously. This air flow contains water vapor to be condensed, but also weakens the radiative cooling of the condenser [38].

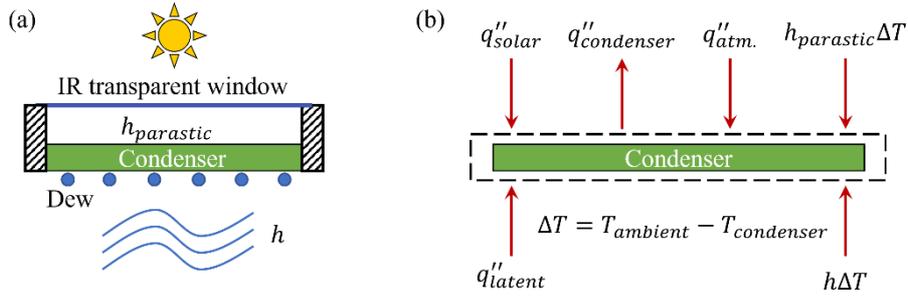

Figure 1: Framework of daytime dew-harvesting. (a) Schematic of the setup for daytime dew-harvesting that dew condenses on the exposed bottom surface instead of the enclosed upper surface, in order not to block the thermal radiation and avoid the sunlight. (b) Energy balance model of daytime dew-harvesting.

We begin by briefly reviewing the framework for nighttime dew-harvesting as reported by Dong, Chen, and co-workers [38]. To simultaneously obtain the temperature of the condenser, $T_{condenser}$, and the mass flux of dew-harvesting, $\dot{m}''$, we couple the energy and the mass transfer between the condenser and its surroundings,

$$q''_{latent} = q''_{condenser}(T_{condenser}) - q''_{atm.}(T_{ambient}, RH) - q''_{conv.}, \quad (1)$$

$$\dot{m}'' = g_{m,H_2O}[\hat{m}_{H_2O,ambient}(T_{ambient}, RH) - \hat{m}_{H_2O,condenser}(T_{condenser}, RH)], \quad (2)$$

where $q_{latent}$ is the latent heat released during the condensation, $q''_{conv.} = (h + h_{parastic})(T_{ambient} - T_{condenser})$ is the convective heat flux, $q''_{condenser}$ and $q''_{atm.}$ are the thermal radiation emitted or absorbed by the condenser [21-23], respectively, all with SI units of W/m². $T_{condenser}$ and $T_{ambient}$ are the temperature of the condenser and the ambient air, respectively. $RH$ is the relative humidity of air. $\dot{m}''$ is the mass flux of dew-harvesting, with units of g/m²-hr. It can be expressed as $\dot{m}'' = 3600 \times q''_{latent}/\Delta$, where $\Delta$ = 2479.5 kJ/kg is the latent heat per unit mass [43]. $g_{m,H_2O}$ is the mass transfer coefficient of the water vapor, and $\hat{m}_{H_2O,ambient}$ and $\hat{m}_{H_2O,condenser}$ are the mass fraction of water vapor in the ambient and near the condenser surface, respectively. Details of these parameters have been discussed in our previous work [38]. Assuming the surrounding air and vapor are ideal gases [44] and applying the strong analogy between the heat and mass transfer [45], Eq. 2 can be simplified to

$$q''_{latent} = (Le)^n \frac{h}{\gamma} [RH \times P_{H_2O}(T_{ambient}) - P_{H_2O}(T_{condenser})], \qquad (3)$$

where $\gamma$ is the psychrometer constant of the surrounding air (67 Pa/K, at 20 °C) [46], $P_{H_2O}$ is the saturation vapor pressure [43], and $Le = 0.87$ is the Lewis number of air at 300 K [45]. As analyzed in Ref. [38], $n$ is different for different scenarios. For example, $n = -2/3$, $-3/4$, or $-1$ for forced external convection, natural convection, or forced internal convection, respectively. All these scenarios lead to $(Le)^n \approx 1.1$ at 300 K.

As a straightforward generalization, we extend the framework above to daytime by subtracting the absorbed solar irradiance, $q''_{solar}$, by the condenser on the right-hand side of Eq. 1 (Fig. 1b),

$$q''_{latent} = q''_{condenser}(T_{condenser}) - q''_{atm.}(T_{ambient}, RH) - q''_{conv.} - q''_{solar}. \qquad (4)$$

Solving Eqs. 2-4, one can simultaneously obtain $T_{condenser}$ and $\dot{m}''$, under any combination of $q''_{solar}$, spectra of the condenser and atmosphere, $RH$, $T_{ambient}$, and convection coefficients, $h$ and $h_{parastic}$.

## 3. Guidelines for photonic designs

In this section, we quantitatively examine effects of the solar reflectivity and the IR emissivity to guide photonic designs for 24-hour day-night dew-harvesting: First, what is the minimum solar reflectivity that is required? Second, what is the quantitative relation between the IR spectral selectivity (the IR emissivity inside vs. outside the atmospheric transparency window, $\varepsilon_{in,8-13}$ vs. $\varepsilon_{out,8-13}$) and the water production yield? In the following, we use Nanjing, a typical city in the east coast of China with humid climates, as a concrete example to address the two questions above, reserving the analysis of a typical city with arid climates to Supplementary Note 1&2.

3.1 Solar reflectivity

To determine the minimum reflectivity of the condenser that is required for effective dew-harvesting under specific solar irradiance, here we fix the IR spectrum (green line in Fig. 2a), but vary the average solar reflectivity $\langle \rho_{solar} \rangle$ from 1 (red line in Fig. 2a) to values of 0.98, 0.95, and 0.92 (black lines). We use the annual average of environmental parameters in Nanjing, including the relative humidity $RH = 75\%$ [48] in Figs. 2a-b and the ambient temperature $T_{ambient} = 16$ °C [49] in Figs. 2a-c. The average peak solar irradiance $G_{solar} = 500$ W/m² in Fig. 2c is estimated using the average global solar radiation $E_{solar} = 13$ MJ/m²-day [50], with a sinusoidal approximation of the solar irradiance in one day-night cycle (Supplementary Note 3). To explore theoretical limits, we follow the analysis of the nighttime dew-harvesting [38] and assume no parasitic convection above the condenser (i.e. $h_{parastic} = 0$) and a convective heat transfer coefficient of $h = 2$ W/m²-K below, which is verified in Supplementary Note 1.

Figure 2b shows the mass flux of dew-harvesting, $\dot{m}''$, of the various spectra in Fig. 2a as a function of the solar irradiance. The red line represents the ideal scenario with perfect solar reflection, corresponding to the nighttime case [38]. It indicates that a minimum $\langle \rho_{solar} \rangle$ of 0.92 is required for 24-hour dew-harvesting in Nanjing. Figure 2c shows the performance of these spectra as a function of $RH$. Dew-harvesting becomes more difficult as $RH$ decreases, even though the atmospheric transparency increases [38,51]. The lower the $RH$, the higher the $\langle \rho_{solar} \rangle$ is required to achieve daytime dew-harvesting. For instance, the minimum $\langle \rho_{solar} \rangle$ is 0.95 for $RH < 40\%$, when assuming $G_{solar} = 500$ W/m². We note that the preceding analyses are based on an ideal IR spectrum (green line in Fig. 2a). For practical IR spectra that deviate from this idealized scenario, the requirement for $\langle \rho_{solar} \rangle$ becomes more stringent.

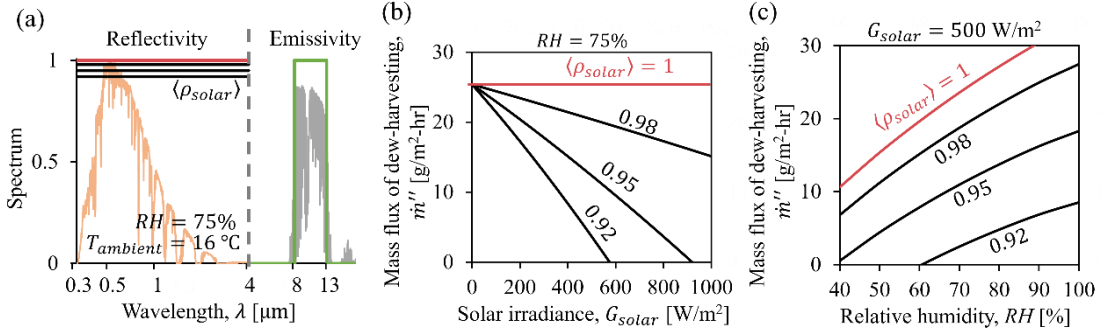

Figure 2: Guidelines for solar reflectivity, $\langle \rho_{solar} \rangle$, to achieve daytime dew-harvesting. (a) Idealized IR spectrum (green) with varied $\langle \rho_{solar} \rangle$: 1 (red), 0.98, 0.95, and 0.92 (black lines). The AM1.5 solar spectrum (orange) and a typical atmospheric transmissivity (gray) are shown as references [47]. (b-c) $\dot{m}''$ as a function of $G_{solar}$, and $RH$, respectively. A minimum $\langle \rho_{solar} \rangle$ of 0.92 is required to achieve 24-hour dew-harvesting in Nanjing. Here we fix $T_{ambient} = 16$ °C, $h_{parastic} = 0$, and $h = 2$ W/m²-K for all calculations.

3.2 Infrared emissivity

We next develop a transfer function that correlates the IR spectral selectivity ($\varepsilon_{in,8-13}$ vs. $\varepsilon_{out,8-13}$) with the water production yield. To simplify the analysis, here we assume zero solar absorptivity, $\langle \rho_{solar} \rangle = 1$. Figure 3a shows three different spectra. Spectrum I is an ideal selective emitter with unity (zero) average emissivity inside (outside) transparency window, $\langle \varepsilon_{in,8-13} \rangle = 1$ ( $\langle \varepsilon_{out,8-13} \rangle = 0$ ). Spectrum II is an ideal IR blackbody with $\langle \varepsilon_{in,8-13} \rangle = \langle \varepsilon_{out,8-13} \rangle = 1$. These two spectra represent typical ideal photonic designs. However, practical designs are less ideal, which can be mimicked using spectrum III with $\langle \varepsilon_{in,8-13} \rangle = 0.9$ and $\langle \varepsilon_{out,8-13} \rangle = 0.1$. The atmospheric spectrum is obtained from ModTran [47]. Following the analysis above, here we also assume $T_{ambient} = 16$ °C, $h_{parastic} = 0$, and $h = 2$ W/m²-K.

Figure 3b shows the resulting temperature of the three spectra, with the dewpoint, $T_{dewpoint}$ (blue dashed), and the fixed $T_{ambient}$ (gray dash-dotted) as references. As $RH$ decreases, the condenser temperatures progressively rise above the $T_{dewpoint}$ threshold, resulting in the cessation of condensation formation. For example, at $RH = 60\%$ [49] and $T_{ambient} = 16$ °C [50], the blackbody spectrum (black; $T_{condenser,(II)}$) cannot cool the condenser below $T_{dewpoint}$, and thus fails to harvest dew, while the practical selective spectrum (red; $T_{condenser,(III)}$) still works. Figure 3c compares the mass flux, $\dot{m}''$, of these three spectra under $T_{ambient} = 16$ °C and $RH = 75\%$ (Nanjing climates). The effect of the spectral selectivity is generalized in Fig. 3d, a contour plot of $\dot{m}''$ as a continuous function of $\langle \varepsilon_{in,8-13} \rangle$ and $\langle \varepsilon_{in,8-13} \rangle$. Figure 3d shows that increasing $\langle \varepsilon_{in,8-13} \rangle$ and decreasing $\langle \varepsilon_{out,8-13} \rangle$ both improve $\dot{m}''$. The slope of the equal-$\dot{m}''$ lines in Fig. 3d is 0.59, indicating that every 10% reduction in $\langle \varepsilon_{out,8-13} \rangle$ is equivalent to an 5.9% increase in $\langle \varepsilon_{in,8-13} \rangle$. In an arid area, e.g. Lanzhou, a typical inland city of China, the slope increases to 0.83, revealing the exacerbated demand for spectral selectivity under low-humidity conditions (Supplementary Note 2).

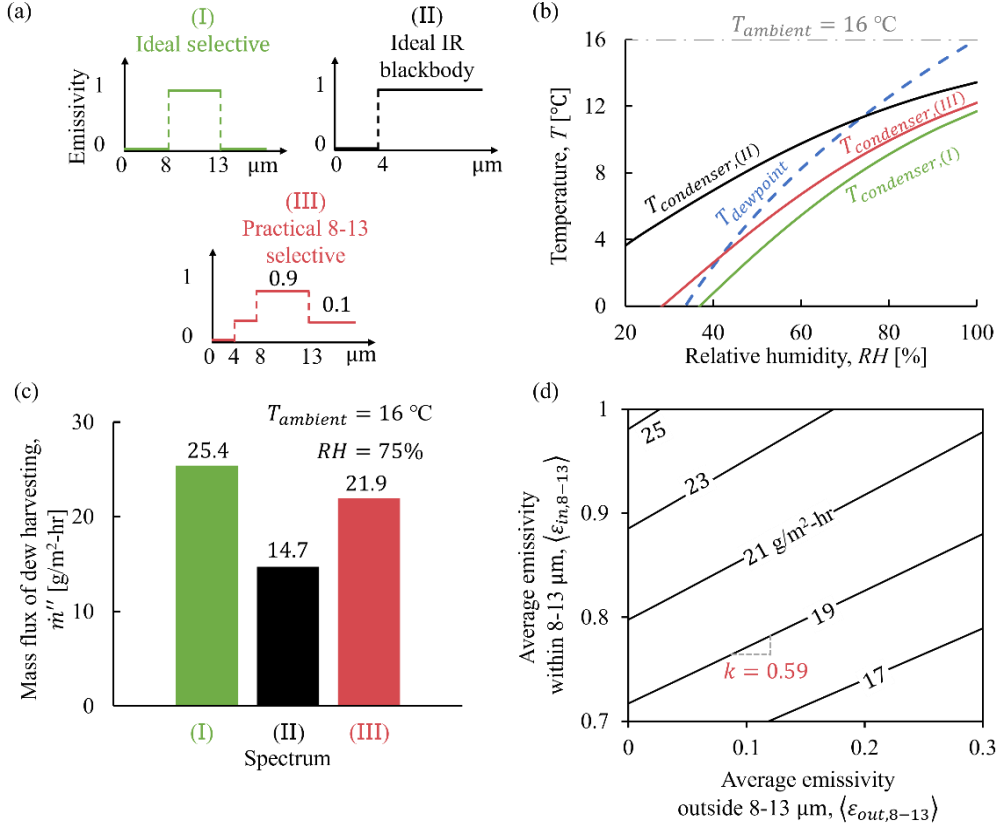

Figure 3: Importance of the spectral selectivity in the IR regime: emissivity inside vs. outside the transparency window, $\langle\varepsilon_{in,8-13}\rangle$ vs. $\langle\varepsilon_{out,8-13}\rangle$. (a) Three spectra, all assuming $\langle\rho_{solar}\rangle = 1$. (b) Calculated temperature of the three spectra (green, black, and red), with the calculated dewpoint temperature ($T_{dewpoint}$, blue) [52] and the fixed ambient temperature ($T_{ambient} = 16$ °C; gray) as references. As $RH$ decreases, $T_{dewpoint}$ (blue) decreases more drastically than the temperature of the condensers, leading to harder and harder condensation. For example, at $RH = 60\%$, the blackbody spectrum (black) fails to harvest dew, while the ideal selective emitter (green) can still cool the condenser below $T_{dewpoint}$. (c) Mass flux, $\dot{m}''$, of the three spectra under Nanjing climates that $T_{ambient} = 16$ °C and $RH = 75\%$. (d) Contour plot of $\dot{m}''$ with respect to $\langle\varepsilon_{in,8-13}\rangle$ and $\langle\varepsilon_{out,8-13}\rangle$ under Nanjing climates, which shows that a 10% decrease in $\langle\varepsilon_{out,8-13}\rangle$ is equivalent to a 5.9% increase in $\langle\varepsilon_{in,8-13}\rangle$. Here we fix $h_{parastic} = 0$, and $h = 2$ W/m²-K.

## 4 A photonic design and virtual experiments

To verify our framework, we provide a multilayer photonic design, consisting of 10 layers (Fig. 4a). A memetic algorithm [53] is used to optimize the spectrum, achieving high solar reflectivity of $\langle\rho_{solar}\rangle = 0.95$, and high IR selectivity of $\langle\varepsilon_{in,8-13}\rangle = 0.92$ and $\langle\varepsilon_{out,8-13}\rangle = 0.37$. Since the photonic design is covered with an IR transparent polyethylene (PE) film (Fig. 1a) [51], hydrolysis of the $MgF_2$ layer is not a concern. In addition, one can replace the PE film periodically to mitigate the degradation of radiative cooling caused by dust accumulation.

A virtual experiment is conducted using measured meteorological datasets in Nanjing, including the solar irradiance, $G_{solar}$ (orange in Fig. 4b), the ambient temperature, $T_{ambient}$ (gray in Fig. 4b), and the relative humidity, $RH$ (Supplementary Note 4). A pronounced diurnal divergence emerges between $T_{ambient}$ and $T_{dewpoint}$ (blue in Fig. 4b), where $\Delta T = T_{ambient} - T_{dewpoint}$ maintains below 2 °C during nighttime but increases to 9.5 °C by noon. This divergence originates from the near-invariant absolute humidity (Supplementary Note 4). For the same reason, the atmospheric spectrum also remains nearly-invariant [51]. We set $h_{parastic} = 0.2$ W/m²-K and $h = 1$ W/m²-K (Fig. 1a), respectively. The former represents a vacuum-based thermal insulation design [23], while the latter ensures dew-harvesting under the peak solar irradiance (Supplementary Note 5).

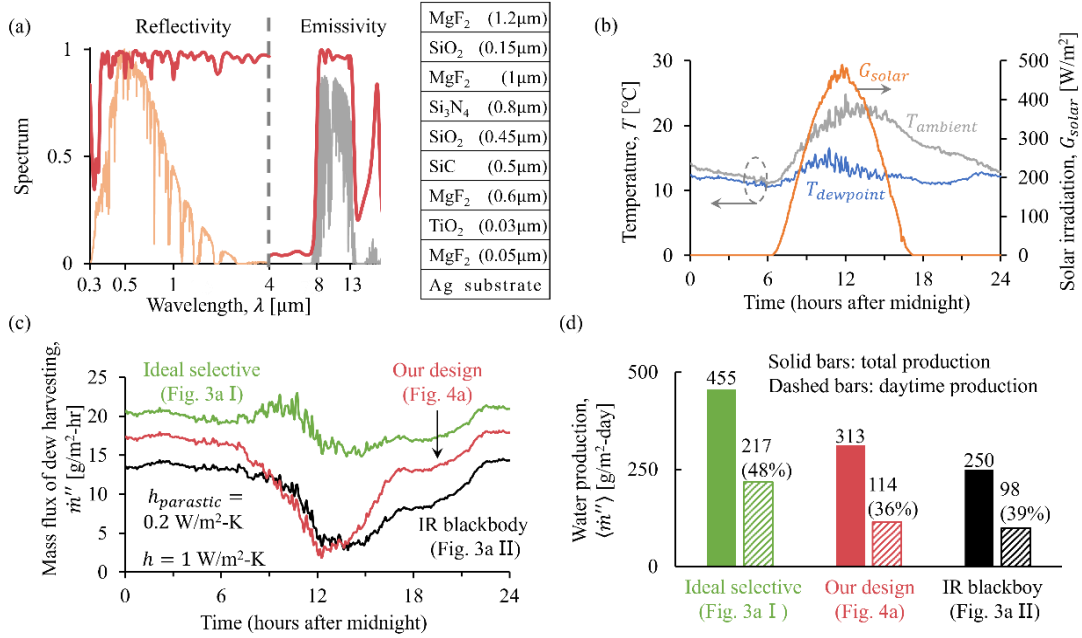

Figure 4: Photonic design and virtual experiment to demonstrate the potential of the 24-hour day-night dew-harvesting. (a) A multilayer photonic design and its spectrum along the normal direction. It has an average reflectivity of $\langle \rho_{solar} \rangle = 0.95$ and averaged emissivity of $\langle \varepsilon_{in,8-13} \rangle = 0.92$ and $\langle \varepsilon_{out,8-13} \rangle = 0.37$. The AM1.5 solar spectrum (orange) and a typical atmospheric transmissivity (gray) are shown as references [47]. (b) Measured meteorological datasets in Nanjing. $T_{dewpoint}$ is calculated based on $T_{ambient}$ and $RH$ [52]. Note the contrast between the nearly-invariant $T_{dewpoint}$ and the drastically varied $T_{ambient}$ throught the cycle. (c-d) Comparison of time-resolved and 24-hour total water production. The more ideal the spectrum, the higher the weighting of the daytime yield.

In Fig. 4c, we calculate the time-resolved dew mass flux $\dot{m}''$ of the three photonic designs (Fig. 3a I&II and Fig. 4a) over a 24-hour period. Our photonic design (red lines in Fig. 4a) can continuously condense water from the surrounding air, even under the peak sunlight at noon. It outperforms the ideal IR blackbody (Fig. 3a II) throughout most of the cycle, except at the peak sunlight when it slightly underperforms due to its non-ideal solar reflectivity. Although neither the ideal selective emitter (Fig. 3a I) nor the IR blackbody (Fig. 3a II) absorbs sunlight, their $\dot{m}''$ decreases around noon due to higher contrast between $T_{dewpoint}$ and $T_{ambient}$ as highlighted in Fig. 4b. The decline is mitigated for the ideal selective emitter due to its stronger cooling ability, demonstrating the advantage of a selective IR spectrum. Figure 4d compares the 24-hour (solid bars) and daytime (from sunrise to sunset; dashed bars) water production $\langle \dot{m}'' \rangle$ (with units of g/m²-day) of the three spectra. The multilayer photonic design achieves nearly 70% of the water production of the ideal selective emitter. More importantly, the daytime water production accounts for about 40% of the total, highlighting the significant potential of daytime dew-harvesting.

## 5 Discussion

In this section, we discuss two practical issues associated with dew-harvesting. First, it is important to distinguish the global versus the local humidity in the application of the framework, otherwise it could lead to unfair comparison between the theory and the experiment. Second, an intuitive trend is that larger total area of the condenser tends to increase the total water production rate, $\dot{m}$ (with units of g/hr).

However, given a total area of the condenser panel, it is not obvious how the layout should be optimized to maximize $\dot{m}$. We address these questions in the following.

5.1 Global vs. local humidity

We compare two configurations that could potentially cause confusions when one compares experimental data to theoretical predictions [40]. In the first configuration (Figs. 5a-b), the water vapor supply is directly from the ambient air, which represents the default setup for dew-harvesting [21-28]. In this case, one has $RH_{global} = RH_{local}$, where $RH_{global}$ ($RH$ in Eq. 1) represents the humidity of the global atmosphere while $RH_{local}$ ($RH$ in Eq. 2) represents the humidity of the local water vapor supply. In the second configuration (Fig. 5c), the ambient air is humidified, e.g. to $RH_{local} = 100\%$, before feeding to the condenser. This setup mimics the scenario that uses radiative cooling to enhance desalination [2,14-20], in which seawater is first evaporated to steam using the sun as the heat source, and subsequently condensed to fresh water using outer space as the heat sink. However, this method is constrained to coastal regions or ocean environments, as it relies on seawater. In this case, one has $RH_{global} < RH_{local}$. As a concrete example, we use condenser in Fig. 4a and assume $G_{solar} = 500$ W/m² and $T_{ambient} = 16$ °C to represent typical climates of Nanjing. We further assume $h = 2$ W/m²-K to maximize the dew-harvesting rate (see details in Fig. 6a).

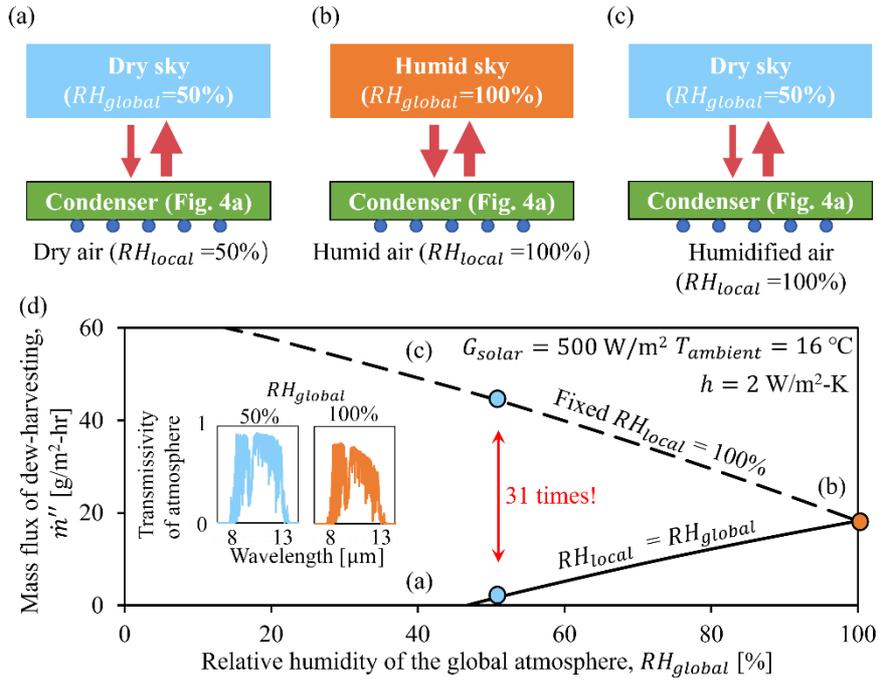

Figure 5: Comparison of two configurations for dew-harvesting. (a-b) A configuration that mimics the common dew-harvesting system, in which the global relative humidity, $RH_{global}$, and the local relative humidity, $RH_{local}$, are the same. (c) A configuration that resembles the desalination system, in which the ambient air is first humidified before feeding to the condenser, and thus $RH_{local}$ is higher than $RH_{globle}$. (d) Dew-harvesting rate of the two configurations. The spectrum design in Fig. 4a is used as the condenser. The three points represent the three scenarios in Fig. 5a-c, respectively.

Figure 5d highlights opposite trends of the two configurations, in which the mass flux of dew-harvesting, $\dot{m}''$, is plotted as a function of $RH_{global}$. For the configuration that resembles the default setup for dew-harvesting (Figs. 5a-b), in which $RH_{global} = RH_{local}$, there are competing effects as $RH_{global}$ increases. On one hand, the sky becomes less transparent (see insets of Fig. 5d: blue vs. orange)

[51], which is detrimental to radiative cooling. On the other hand, higher $RH_{global}$ (and thus higher $RH_{local}$ in this scenario) means more water vapor in the surrounding air, which is beneficial to the dew-harvesting. The latter wins the competition, and thus $\dot{m}''$ and $RH_{globle}$ are positively correlated (black solid line). In contrast, for the configuration that mimics the desalination (Fig. 5c), in which $RH_{local}$ is fixed at 100%, the increase of $RH_{globle}$ weakens radiative cooling, and therefore $\dot{m}''$ and $RH_{globle}$ are negatively correlated (black dashed line). These opposite trends of the two configurations emphasize the importance of choosing the correct model if one wants to compare their experiments to the theoretical predictions [40].

5.2. Centralized vs. distributed layout of condensers

We address the optimization problem of the layout of condensers based on the panel-size-dependent convection coefficient, $h$, which is not well appreciated by the community of radiative cooling. In the following, we will utilize one of the major conclusions from Ref. [38] that $h$ can be optimized to maximize the mass flux when $RH < 100\%$, and this optimized $h_{opt.}$ corresponds to an optimized size of the condenser panel.

On one hand, the correlation between the length, $L$, of the condenser (with SI units of m) and the resulting $h$ is well documented in heat transfer textbooks under various scenarios (e.g. Ref. [45,54]). As a concrete example, here we use the forced external laminar convection to illustrate the dependence, reserving the complicated multi-mode practical scenario to Supplementary Note 6. In this simplified case, the dependence is cast into correlations among three dimensionless groups [45,54],

$$Nu_L = 0.664 Re^{1/2} Pr^{1/3}, \qquad (5)$$

where the Nusselt number, $Nu_L = hL/k$ is the dimensionless heat transfer coefficient, the Reynolds number, $Re = v_{wind}L/\nu$, indicates the competition between the inertial force and the viscous force, and the Prandtl number, $Pr = \nu/\alpha$, reflects the relative strength between the kinetic and the thermal diffusivities. Note here, $k$ is the thermal conductivity of air (with SI units of W/m-K), $\nu$ is the kinetic viscosity of air (with SI units of m²/s), $\alpha$ is the thermal diffusivity of air (with SI units of m²/s) [45,54]. We have $Pr = 0.71$ for air near 300 K under the standard atmospheric pressure, satisfying the criterion to apply Eq. 5, $Pr \geq 0.6$ [45,54]. From Eq. 5, one obtains the correlation between the size of the condenser and the convection coefficient,

$$h \propto L^{-1/2}. \qquad (6)$$

We emphasize that the simple correlation described by Eqs. 5-6 are valid for forced external laminar convection of air near 300 K under the standard atmospheric pressure. For the practical scenario described by Fig. 1a, the correlation is much more complicated (black line in Fig. 6a; see details in Supplementary Note 6).

On the other hand, there is an optimized convection coefficient, $h_{opt.}$, to maximize the mass flux, $\dot{m}''$ (with units of g/m²-hr¹), due to the competition between increasing the water-vapor supply and weakening radiative cooling as $h$ increases [38]. For example, using the photonic design in Fig. 4a, $h_{opt.}$ for the typical climate in Nanjing is 2.0 W/m²-K, as indicated in Fig. 6a. The correlation described in Fig. S5 of the Supplementary Note 6 links this $h_{opt.}$ to an optimized size of the condenser, $L_{opt.}$.

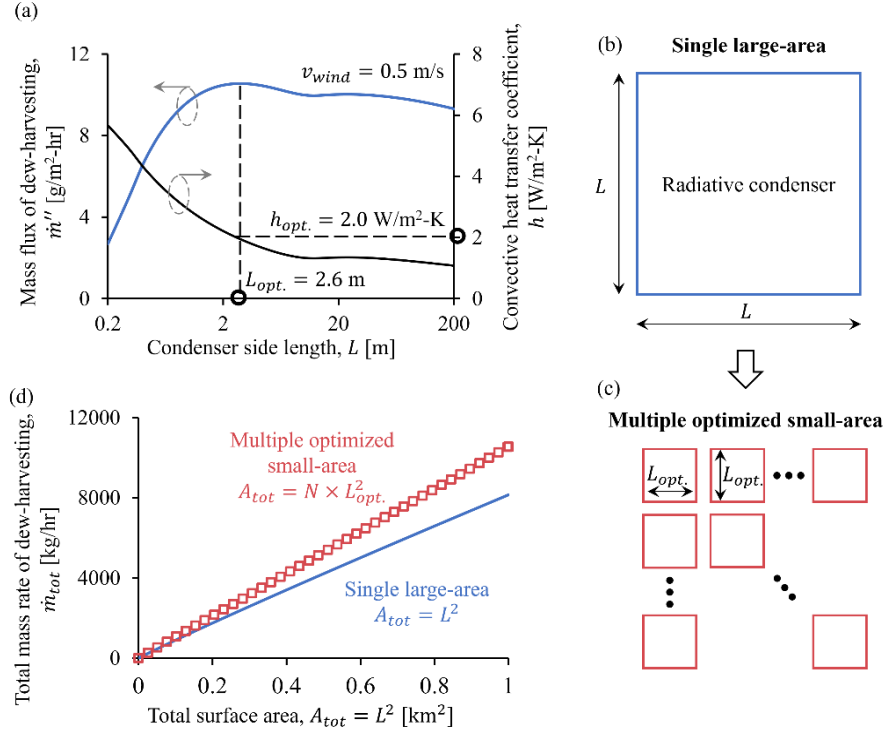

Figure 6: Condenser layout: centralized vs. distributed. (a) To maximize the mass flux, $\dot{m}''$ (with units of g/m²-hr), there is an optimized convection coefficient, $h_{opt.}$, and therefore a corresponding optimized size of the condenser, $L_{opt.}$. For example, using the spectrum in Fig. 4a and typical climates in Nanjing, e.g. $T_{ambient} = 16$ °C, $RH = 75\%$, $G_{solar} = 500$ W/m², and $v_{wind} = 0.5$ m/s, one obtains $h_{opt.} = 2.0$ W/m²-K, and $L_{opt.} = 2.6$ m. (b-d) Comparison of two configurations, which have the same total surface area, $A_{tot} = L^2 = N \times L_{opt.}^2$. As a concrete example, for a set of parameters, $N = 39625$, $L_{opt.} = 2.6$ m, $L = 1000$ m, and thus $A_{tot} = 1$ km², the distributed multiple small-area layout (red) enhances the dew yield by 30% as compared to the centralized single large-area layout (blue).

Figure 6d compares the mass flux, $\dot{m}''_{tot}$ (with units of kg/hr), of two configurations, a single large-area condenser panel (blue; Fig. 6b) and multiple small-area panels (red; Fig. 6c). We note that these two configurations have the same total surface area of condensers $A_{tot} = L^2 = N \times L_{opt.}^2$, which is designated as the x-axis. To illustrates the optimization process, we first obtain the optimized convection coefficient, $h_{opt.} = 2.0$ W/m²-K, for the typical climate in Nanjing (Fig. 6a; see details in Supplementary Note 1). The corresponding optimized length of the unit cell, $L_{opt.} = 2.6$ m, can be obtained using the correlations in Supplementary Note 6. Note here, we assume $v_{wind} = 0.5$ m/s in the calculations, which can be achieved by applying wind shields [55]. This optimized unit cell corresponds to the maximized mass flux, $\dot{m}''_{max} = 10.5$ g/m²-hr, under this specific climate condition. Next, we use these unit cells as building blocks to assemble condenser arrays. Analogous to the concentrating solar plants [56], here we set $A_{tot} = 1$ km² with $L = 1$ km (Fig. 6b). Correspondingly, the distributed array (Fig. 6c) has $N = 39625$ unit-cells with an optimized size of $L_{opt.} = 2.6$ m. This distributed layout gives a total mass rate of $\dot{m}_{tot,max} = A_{tot} \times \dot{m}''_{max} = 10500$ kg/hr. In contrast, a single large-area panel with the same total area $A_{tot}$ leads to a convection coefficient of $h = 0.81$ W/m²-K and a corresponding $\dot{m}'' = 8.1$ g/m²-hr. As a result, this configuration (blue in Fig. 6d) gives $\dot{m}_{tot} = A_{tot} \times \dot{m}'' = 8100$ kg/hr, which is 30% less than the optimized distributed configuration (red in Fig. 6d). From a different point of view, to harvest dew at the same mass rate of $\dot{m}_{tot} = 8100$ kg/hr, the distributed configuration saves 23% surface areas as compared to the single large-area one.

# 6 Conclusion

In summary, we developed a framework to predict the yield of the dew-harvesting round-the-clock. Using Nanjing as an example, we outlined guidelines for photonic designs to achieve 24-h day-night dew-harvesting, even under the peak of the solar irradiance. In the solar spectrum, a minimum average reflectivity of 0.92 is required for $G_{solar} = 500$ W/m$^2$. In the IR spectrum, selectivity is the key. For example, every 10% reduction of the emissivity outside the transparency window (i.e. 8-13 μm) is equivalent to a 5.9% (8.3%) enhancement of that inside the window in Nanjing (Lanzhou). To validate these guidelines, we provided a numerical example of the selective photonic design, consisting of 10 layers. With this design, we conducted a virtual experiment to achieve continuous day-night dew-harvesting, in which 40% of the water yield is contributed by daytime. This demonstration underlines the potential of dew-harvesting at daytime. We ended by commenting on the real-world applications of this framework, and highlighted the importance of the consistency between experiments and models, as well as the layout optimization of condensers. We anticipate this framework can also be used in other applications such as estimating the loss of water droplets due to evaporation in the liquid droplet radiators for heat rejection in space [57-60].

**Conflict of interest statement**: The authors declare no conflicts of interest regarding this article.

**Research funding**: This work was supported by the National Natural Science Foundation of China (52376051) and Postgraduate Research & Practice Innovation Program of Jiangsu Province (KYCX22_0197).

# Photonic Structures to Achieve High-Performance Dew-Harvesting in a 24-h Day-Night Cycle

ZHANG Zheng[1], DONG Lining[2], DONG Minghao[1], RUAN Shiting[2], & CHEN Zhen[1*]

[1] *Jiangsu Key Laboratory for Design and Manufacturing of Precision Medicine Equipment, School of Mechanical Engineering, Southeast University, Nanjing 210096, China*

[2] *Shanghai Institute of Satellite Engineering, Shanghai 201109, China*

\* *To whom correspondence should be addressed: zhenchen@seu.edu.cn*

## Supplementary Notes

## 1  Competition between water-vapor supply and radiative cooling: an optimized $h$

There is an optimized convective heat transfer coefficient $h_{opt.}$ for scenarios with $RH < 100\%$ [1]. The trade-off is that a lower $h$ enhances the radiative cooling process, while a higher $h$ facilitates the supply of water vapor. As shown in Fig. S1a, a lower $\langle \rho_{solar} \rangle$ makes cooling more challenging, requiring a lower $h$ to compensate the temperature reduction.

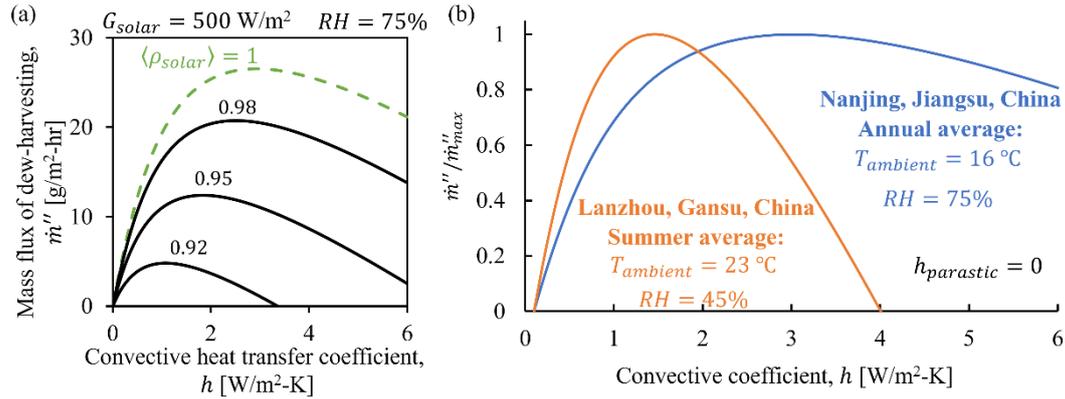

Figure S1: Optimization of $h$. (a) Optimized $h$ vs. solar reflectivity. Green dashed and black solid lines represent the ideal IR selective emitter with $\langle \rho_{solar} \rangle = 1$, 0.98, 0.95 and 0.92, respectively. Environmental parameters: $T_{ambient} = 16°C$, $G_{solar} = 500$ W/m² and $RH = 75\%$ (Nanjing, a typical east-coast city of China). (b) Optimized $h$ vs. climates. Lanzhou (orange line) has a lower $h_{opt.}$ than that of Nanjing (blue line).

We extend our analysis to Lanzhou, a typical inland city of China with dry climate. As discussed above, worse cooling ability necessitates a lower $h_{opt.}$ to compensate for the cooling requirement (Fig. S1a). Lanzhou, with an average absolute humidity $AH = 9.26$ g/m³ (see Supplementary Note 4), has a more transparent sky [2] and enables stronger radiative cooling than Nanjing ($AH = 10.23$ g/m³, see supplementary 5). However, Fig. S1b indicates that Lanzhou requires a lower $h_{opt.}$ compared to Nanjing. This is because, in Lanzhou, the dew point temperature is calculated as $T_{dp,Lanzhou} = 10.4\ °C$, while $T_{dp,Nanjing} = 11.6\ °C$ [3]. The condenser needs to be cooled by $\Delta T_{Lanzhou} = T_{ambient,Lanzhou} - T_{dp,Lanzhou} = 12.6\ °C$ in Lanzhou, whereas in Nanjing, $\Delta T_{Nanjing} = 4.4\ °C$ suffices. Therefore, lower $h$ is needed to meet the larger cooling demand.

## 2 Optimization of the IR emissivity in arid areas

Here we perform a similar analysis of Fig. 3d for Lanzhou, a typical inland city of China in arid area, with an annual average $RH = 45\%$ [4] and $T_{ambient} = 23°C$ [5] in summer. Figure S2 shows the same conclusion as Fig. 3d: increasing $\langle \varepsilon_{in,8-13} \rangle$ and decreasing $\langle \varepsilon_{out,8-13} \rangle$ both improve dew-harvesting performance. The slope of the equal-mass-flux lines in Fig. S3 is 0.83, indicating that each 10% reduction in $\langle \varepsilon_{out,8-13} \rangle$ contributes equivalently to dew-harvesting as an 8.3% increase in $\langle \varepsilon_{in,8-13} \rangle$. Comparing the slope of 0.59 in Nanjing, this larger slope highlights that the selective IR emissivity is more important in arid areas.

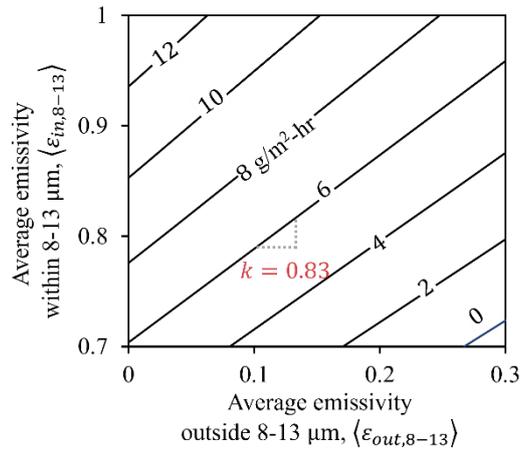

Fig. S2: Contour plot of the mass flux of dew-harvesting in Lanzhou, with respect to average emissivity within and outside 8-13μm.

## 3 Modeling the solar irradiance in a day-night cycle

We approximate the solar irradiance to vary with time in a half-sinusoidal form:

$$q''_{solar} = \begin{cases} G_{solar} \sin(2\pi t/\tau + \phi_{solar}), & day \\ 0, & night \end{cases}, \tag{S1}$$

where $t$ is the time and $\tau$ is the diurnal period with SI units of s. Here the phase $\phi_{solar}$ is chosen so that $2\pi t/\tau + \phi_{solar} = 0$ at sunrise. Therefore, the annual average value of the daily normal solar radiation can be expressed as

$$E_{solar} = \int_0^\tau q''_{solar} dt = \frac{\tau G_{solar}}{\pi}. \tag{S2}$$

According to Ref. 6, the $E_{solar}$ of Nanjing is measured of 13 MJ/m²-day, and thus we can estimate $G_{solar}$ as 500 W/m².

## 4 Absolute humidity

Absolute humidity $AH$ represents the total water content per unit volume of air (with SI units of g/m³), which directly affects the atmospheric transmissivity [2]. The absolute humidity can be calculated as [8]

$$AH = \frac{0.217 \times P_{vapor}}{T_{ambient}}, \quad (S3)$$

where $P_{vapor}$ is the vapor pressure with SI units of Pa, and $T_{ambient}$ is the ambient temperature with SI units of K. $P_{vapor}$ can be calculated as [7]

$$P_{vapor} = 6.1078 \times 10^{\frac{7.5 \times (T_{ambient} - 273.15)}{T_{ambient} - 35.85}} \times RH. \quad (S4)$$

Based on the measured $T_{ambient}$ (gray line in Fig. 4b in the main text) and $RH$ (blue line in Fig. S3), the absolute humidity $AH$ is obtained (black line in Fig. S3). It remains nearly constant during a day-night cycle with an average value of 10.6 g/m³.

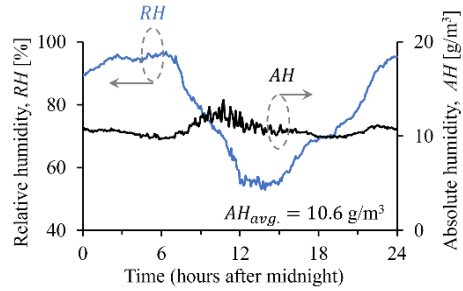

Figure S3: Measured $RH$ and calculated $AH$.

## 5 Effects of convection on the virtual experiment

In the virtual experiment, with a setup as in Fig. S4a and a photonic design as in Fig. S4b (the same as in Fig. 4a of the main text), we assume $h_{parastic} = 0.2$ W/m²-K above the condenser, representing a practical vacuum chamber [9]. Setting $h = 1$ W/m²-K for convection below (black line) allows for effective 24-hour dew-harvesting even under direct sunlight. When $h$ is increased to 2 W/m²-K (red line), it becomes more challenging to cool the condenser below the dew point around noon, hindering dew harvesting over the full 24-hour period. However, this low $h$ scenario is advantageous under weak solar irradiance.

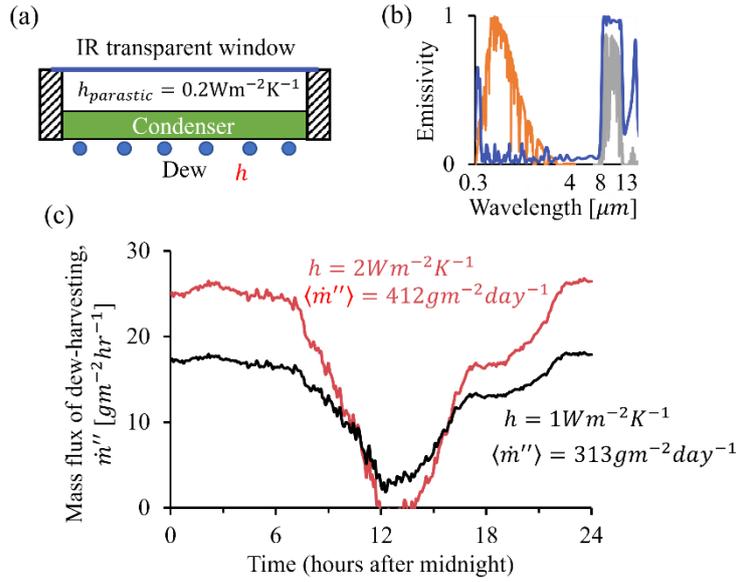

Figure S4: Effects of $h$ on the virtual experiment. (a-b) Illustration of the experimental setup and the photonic design used in the virtual experiment. (c) Real-time water production of the photonic design. The black line represents the scenario of $h = 1$ W/m²-K, which ensures dew-harvesting in a day-night cycle with a total daily yield of 313 g/m²-day. The red line represents the scenario of $h = 2$ W/m²-K, which fails in harvesting dew at noon, but results in a higher total daily yield of 412 g/m²-day.

# 6  The effective convection coefficient in Fig. 6 of the main text

Modeling the heat convection of the external air flow through a horizontal plate (Fig. 6) in real applications is a bit more complicated, since it involves multi-modes. First, both the natural and the forced convection should be considered, since the cold condenser is on top of the setup (Fig. 1a) and the criterion, $Gr/Re^2 \sim 1$, where $Gr$ is the Grashof number, and $Re$ is the Reynolds number, as will be discussed in detail in the following. Second, both the laminar and the turbulent flow should be taken into account in the analysis of the forced convection, since the local $Re$ spans a wide range. These complexities lead to an effective global Nusselt number [10,11]

$$Nu = \left(Nu_n^3 + Nu_f^3\right)^{1/3}, \tag{S5}$$

where $Nu_n$ and $Nu_f$ are the global Nusselt numbers of the natural and the forced convection.

We first obtain $Nu_f$. One usually defines a local Reynolds number, $Re_x = v_{wind} x/\nu$, to determine whether the flow is laminar, turbulent, or transitioning between the two modes, where $x$ is the distance from the leading edge, $v_{wind}$ is the wind speed, and $\nu$ is the kinetic viscosity. Combining the local Nusselt number, $Nu_{x,f} = h_{x,f} x/k$, and the Prandtl number, $Pr$, one can calculate the forced convection coefficient $h_{x,f}$ at each location, i.e. the local convection coefficient.

Depending on the local Reynolds number, $Re_x = v_{wind} x/\nu$, the air flow is either laminar with a local Nusselt numbers [10]

$$Nu_{x,lam.} = 0.332 Re_x^{1/2} Pr^{1/3}, (Re_x \leq 2 \times 10^5) \tag{S6}$$

turbulent with

$$Nu_{x,turb.} = 0.0296 Re_x^{4/5} Pr^{3/5}, (Re_x \geq 5 \times 10^5) \tag{S7}$$

or, transitioning between the two modes with

$$Nu_{x,trans.} = Nu_{x,lam.}(Re_{crit.}) \times \left(\frac{Re_x}{Re_{crit.}}\right)^c, (2 \times 10^5 < Re_x < 5 \times 10^5) \tag{S8}$$

where $c = 0.9922 \times \log_{10} Re_{crit.} - 3.013$, and $Re_{crit.} = 2 \times 10^5$.

The local forced convection coefficient, $h_{x,f}$, and the global one, $h_f$, can then be calculated as [9]

$$h_{x,f}(x) = \frac{Nu_x k}{x}, \tag{S9}$$

$$h_f = \frac{1}{L} \int_0^L h_x(x) dx. \tag{S10}$$

Figure S5 shows $Nu_{x,f}$ and $h_{x,f}$ of a condenser with $L = 30$ m under $v_{wind} = 0.5$ m/s. In addition, we assume the temperature of air is around 15°C, and therefore the kinetic viscosity $\nu = 1.48 \times 10^{-5}$ m²s, the thermal conductivity $k = 0.025$ W/m-K, and the Prandtl number $Pr = 0.71$ [8].

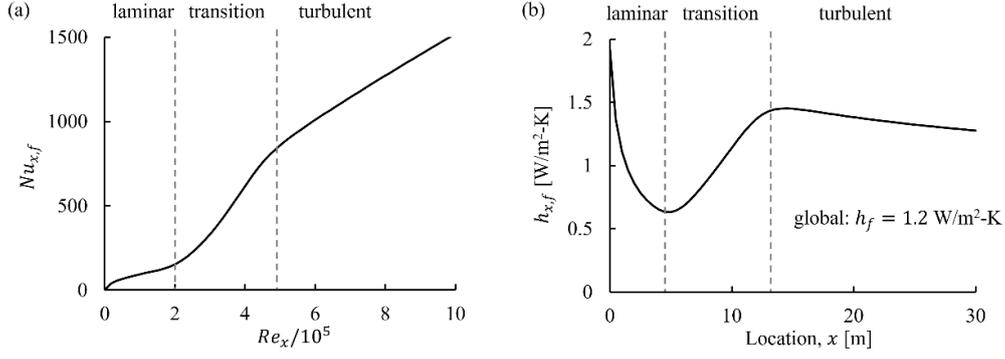

Figure S5: Estimation of the local forced convection coefficient. (a) $Nu_{x,f}$ as a function of $Re_x$. (b) $h_{x,f}$ as a function of locations. $h_{x,f} \propto x^{-1/2}$ and $h_{x,f} \propto x^{-1/5}$ in the laminar and the turbulent region, respectively. In the transitioning region, $h_{x,f}$ increases rapidly as $h_{x,f} \propto x^{1.7}$. The global convection coefficient, $h_f = 1.2$ W/m²-K is obtained by averaging over the length of the condenser.

We next compute $Nu_n$. In the setup (Fig. 1a of the main text), the warm air flows beneath the cold condenser. In addition, the criterion, $Gr/Re^2 = \frac{g\alpha_v \Delta T L}{v_{wind}^2}$, In this case, the natural convection cannot be neglected. The global Nusselt number of natural convection can be calculated using [12].

$$Nu_n = \begin{cases} 0.54(GrPr)^{1/4}, & 10^4 \leq GrPr \leq 10^7 \\ 0.15(GrPr)^{1/4}, & 10^7 \leq GrPr \leq 10^{11} \end{cases}, \quad \text{(S11)}$$

where $Gr = g\alpha_v \Delta T L^3 / v^2$ is the Grashof number. Here $g = 9.8$ m/s² is the acceleration of gravity, $\alpha_v = 1/T = 2/(T_{ambient} + T_{condenser})$ is the coefficient of volume change (assuming ideal gas), $\Delta T = T_{ambient} - T_{condenser}$ is the temperature difference. According to the calculations in the main text, we set $T_{ambient} = 16°C$ and $T_{condenser} = 9°C$.

With $Nu_n$ and $Nu_f$ obtained, one can finally compute the effective global Nusselt number using Eq. S5, and the corresponding effective convection coefficient,

$$h = \frac{Nu k}{L}, \quad \text{(S12)}$$

which is shown in Fig. 6a of the main text.